# Tests for partial correlation between repeatedly observed nonstationary nonlinear timeseries

Kenneth D. Harris[1*], Alex E. Yuan[2*]
[1]UCL Institute of Neurology, Queen Square, London WC1N 3BG, UK. * Correspondence: kenneth.harris@ucl.ac.uk, alexericyuan@gmail.com.
[2]Independent Researcher, Seattle WA, USA.

**We describe two families of statistical tests to detect partial correlation in vectorial timeseries. The tests measure whether an observed timeseries Y can be predicted from a second series X, even after accounting for a third series Z which may correlate with X. They do not make any assumptions on the nature of these timeseries, such as stationarity or linearity, but they do require that multiple statistically independent recordings of the 3 series are available. Intuitively, the tests work by asking if the series Y recorded on one experiment can be better predicted from X recorded on the same experiment than on a different experiment, after accounting for the prediction from Z recorded on both experiments.**

IN many fields of science, observations consist of timeseries. These timeseries are often autocorrelated, meaning that observations from a single timeseries at different times are not statistically independent. Such autocorrelations mean that standard statistical tests, which make an assumption of independent, identically distributed observations, cannot be applied [1–5]. Common approaches to this problem include filtering to remove autocorrelations[3]; corrections to standard tests such as the $t$-test for Pearson correlation[6–8]; and synthesis of surrogate data[9]. However all these approaches make assumptions such as stationarity or linearity the timeseries in question, which are often untrue or impossible to show in practice.

This problem is greatly ameliorated if one has access to multiple repetitions of the experiment. For example, consider an experiment yielding two timeseries $\boldsymbol{X}$ and $\boldsymbol{Y}$ which we consider as vectors of length $T$. The timeseries are statistically independent if $\mathbb{P}(\boldsymbol{X},\boldsymbol{Y}) = \mathbb{P}(\boldsymbol{X})\mathbb{P}(\boldsymbol{Y})$: in other words, if there is no correlation between the entire history of $\boldsymbol{X}$ and the entire history of $\boldsymbol{Y}$. Given only a single observation of the vectors $\boldsymbol{X}$ and $\boldsymbol{Y}$, we cannot test for violation of independence without making further assumptions: methods such as standard tests for Pearson correlation, which assume independence of timepoints, give erroneously significant "nonsense correlations"[5], and corrections to these make further assumptions such as stationarity. However, if the experiment is repeated $N$ times, and we assume that the histories $\boldsymbol{X}_i$ and $\boldsymbol{Y}_i$ $(i = 1 \dots N)$ observed on each repeat are independent vectors, we can employ a "session permutation" method to test for independence [10], without any further assumptions. This method essentially asks whether the relationship between $\boldsymbol{X}_i$ and $\boldsymbol{Y}_i$ is larger than that between $\boldsymbol{X}_i$ and $\boldsymbol{Y}_j$ for $i \neq j$, using a permutation test that randomizes across experiments. The test can be based on any measure of similarity between timeseries, including but not limited to Pearson correlation.

Here, we describe two families of tests for partial correlation between repeatedly observed timeseries. We assume that each of $N$ experiments yielded 3 vectorial timeseries $\boldsymbol{X}, \boldsymbol{Y}$ and $\boldsymbol{Z}$. We ask whether a correlation between $\boldsymbol{X}$ and $\boldsymbol{Y}$ exists beyond a common effect of $\boldsymbol{Z}$. We do not assume that time series are stationary or linear, in fact $\boldsymbol{X}, \boldsymbol{Y}$ and $\boldsymbol{Z}$ can be arbitrary vectors rather than timeseries.

## The problem

We assume $N$ statistically independent experiments, each of which gives an observation of three vectorial timeseries: $\boldsymbol{X}_i$, $\boldsymbol{Y}_i$ and $\boldsymbol{Z}_i$, for $i = 1 \dots N$, considered as matrices of sizes $T \times p_i$, $T \times q_i$ and $T \times r_i$, respectively. Write $\boldsymbol{\mathcal{X}}$ for the collection of all $\boldsymbol{X}_i$ and $\boldsymbol{Z}_i$. We would like to test the null hypothesis that

$$\boldsymbol{Y}_i = \boldsymbol{Z}_i \boldsymbol{W}_i + \boldsymbol{E}_i$$

where the $\boldsymbol{E}_i$ are timeseries independent of each other and of $\boldsymbol{\mathcal{X}}$, and $\boldsymbol{W}_i$ is an unknown deterministic $p_i \times r_i$ matrix. This null hypothesis models the idea that $\boldsymbol{Y}$ may be linearly predicted from $\boldsymbol{Z}$ but there is no further effect of $\boldsymbol{X}$ on $\boldsymbol{Y}$ once the effect of $\boldsymbol{Z}$ is considered, even though $\boldsymbol{X}$ and $\boldsymbol{Z}$ may themselves be correlated.

## Test 1: An approximate test using orthogonal projections

For a $T \times n$ matrix $\boldsymbol{A}$, let $\mathcal{M}$ be the function that maps $\boldsymbol{A}$ to the $T \times T$ matrix performing the orthogonal projection onto the columns of $\boldsymbol{A}$. Specifically, one can take $\mathcal{M}(\boldsymbol{A}) = \boldsymbol{U}\boldsymbol{U}^T$, where $\boldsymbol{U}\boldsymbol{\Sigma}\boldsymbol{V}^T$ is the compact singular value decomposition of $\boldsymbol{A}$ (Ref. [11], p. 82).

Let $\boldsymbol{P}_i = \boldsymbol{I} - \mathcal{M}(\boldsymbol{Z}_i)$, so $\boldsymbol{P}_i$ is the $T \times T$ projection matrix orthogonal to all columns of $\boldsymbol{Z}_i$. Note that if the null hypothesis is true, $\boldsymbol{P}_i\boldsymbol{Y}_i = \boldsymbol{P}_i\boldsymbol{E}_i$. Let $\rho(\boldsymbol{X};\ \boldsymbol{Y})$ be a measure of the predictability of a vector timeseries $\boldsymbol{Y}$ from a second timeseries $\boldsymbol{X}$, for example the Pearson coefficient of multiple linear regression. For a statistical test, you might first think to ask if $\rho(\boldsymbol{X}_i;\ \boldsymbol{P}_i\boldsymbol{Y}_i)$ is bigger than $\rho(\boldsymbol{X}_j;\ \boldsymbol{P}_i\boldsymbol{Y}_i)$ for $i \neq j$, i.e. if the relationship of $\boldsymbol{E}$ and $\boldsymbol{X}$ is strongest when they come from the same experiment. But $\rho(\boldsymbol{X}_i;\ \boldsymbol{P}_i\boldsymbol{Y}_i) = \rho(\boldsymbol{X}_i;\ \boldsymbol{P}_i\boldsymbol{E}_i)$ is not identically distributed to $\rho(\boldsymbol{X}_j;\ \boldsymbol{P}_i\boldsymbol{Y}_i) = \rho(\boldsymbol{X}_j;\ \boldsymbol{P}_i\boldsymbol{E}_i)$ for $j \neq i$: even though $\boldsymbol{E}_i$ is independent of both $\boldsymbol{X}_i$ and $\boldsymbol{X}_j$, $\boldsymbol{X}_i$ and $\boldsymbol{P}_i$ can be dependent under the null hypothesis but $\boldsymbol{X}_j$ and $\boldsymbol{P}_i$ cannot.

Instead, let $\boldsymbol{P}_{i,j} = \boldsymbol{I} - \mathcal{M}([\boldsymbol{Z}_i\ \boldsymbol{Z}_j])$ where $[\boldsymbol{Z}_i\ \boldsymbol{Z}_j]$ is a horizontal concatenation; thus $\boldsymbol{P}_{i,j}$ is the $T \times T$ projection matrix orthogonal to the columns of both $\boldsymbol{Z}_i$ and $\boldsymbol{Z}_j$. Note that $\boldsymbol{P}_{j,i} = \boldsymbol{P}_{i,j}$, because the orthogonal projection onto a given subspace is a unique matrix (Ref. [11], p. 82). We will compare $\rho(\boldsymbol{X}_i;\ \boldsymbol{P}_{i,j}\boldsymbol{Y}_i)$ to $\rho(\boldsymbol{X}_j;\ \boldsymbol{P}_{i,j}\boldsymbol{Y}_i)$, Specifically, define

$$G_i = \frac{1}{N}\sum_{j=1}^{N} \rho(\boldsymbol{X}_i;\ \boldsymbol{P}_{i,j}\boldsymbol{Y}_i) - \rho(\boldsymbol{X}_j;\ \boldsymbol{P}_{i,j}\boldsymbol{Y}_i)$$

Since $\boldsymbol{P}_{i,j}\boldsymbol{Y}_i = \boldsymbol{P}_{i,j}\boldsymbol{E}_i$, it follows that $G_i = \frac{1}{N}\sum_{j=1}^{N} \rho(\boldsymbol{X}_i;\ \boldsymbol{P}_{i,j}\boldsymbol{E}_i) - \rho(\boldsymbol{X}_j;\ \boldsymbol{P}_{i,j}\boldsymbol{E}_i)$. Under the null hypothesis, the $\boldsymbol{E}_i$ are independent, and thus the $G_i$ are conditionally independent given $\boldsymbol{\mathcal{X}}$, although they need not be identically distributed.

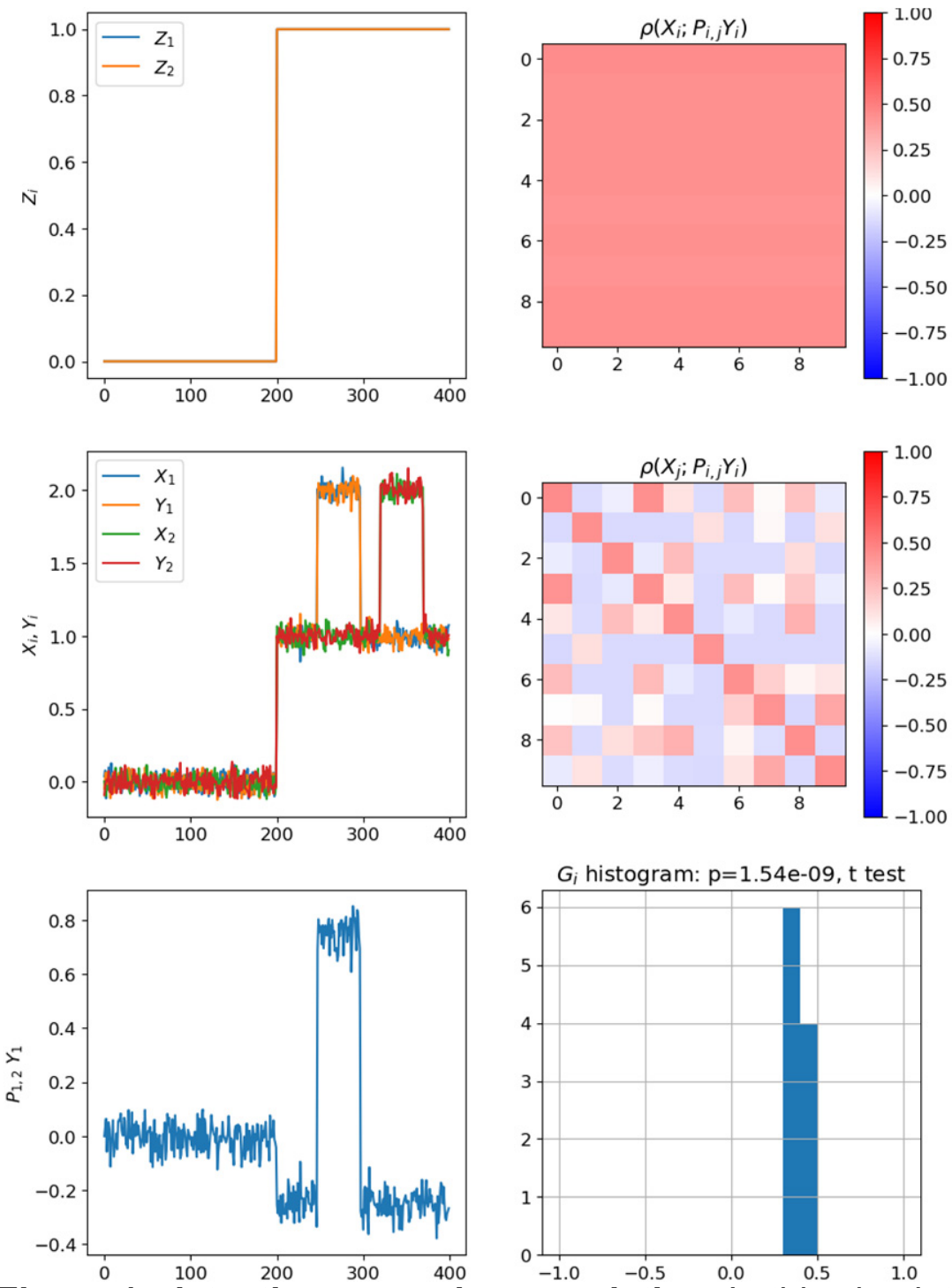

**Figure 1: detecting a genuine correlation.** In this simulation $\boldsymbol{X}_i$ and $\boldsymbol{Y}_i$ are both equal to a sum of two step functions, plus independent additive noise. One step function is constant across experiments. $\boldsymbol{Z}_i$ equals the constant step function, and after removing it from $\boldsymbol{Y}_i$, the residual $\boldsymbol{P}_{i,j}\boldsymbol{Y}_i$ is reliably more correlated to $\boldsymbol{X}_i$ than to $\boldsymbol{X}_j$, and statistical significance is detected.

Under the null, the expectation of $\sum_{i=1}^{N} G_i$ conditional on $\boldsymbol{\mathcal{X}}$ is zero. To see this, first use the fact that $\boldsymbol{P}_{j,i} = \boldsymbol{P}_{i,j}$ to rewrite $\sum_{i=1}^{N} G_i$ as follows:

$$\begin{aligned}\sum_{i=1}^{N} G_i &= \frac{1}{N}\sum_{i,j=1}^{N} \rho(\boldsymbol{X}_i;\ \boldsymbol{P}_{i,j}\boldsymbol{E}_i) - \rho(\boldsymbol{X}_j;\ \boldsymbol{P}_{i,j}\boldsymbol{E}_i) \\ &= \frac{1}{N}\sum_{i,j=1}^{N} \rho(\boldsymbol{X}_i;\ \boldsymbol{P}_{i,j}\boldsymbol{E}_i) - \rho(\boldsymbol{X}_i;\ \boldsymbol{P}_{i,j}\boldsymbol{E}_j)\end{aligned}$$

Next, because $\boldsymbol{E}_i$ and $\boldsymbol{E}_j$ are identically distributed and independent of everything else,

$$\begin{aligned}&\mathbb{E}[\textstyle\sum_{i=1}^{N} G_i\,|\boldsymbol{\mathcal{X}}] \\ &= \frac{1}{N}\sum_{i,j=1}^{N} \mathbb{E}[\rho(\boldsymbol{X}_i;\ \boldsymbol{P}_{i,j}\boldsymbol{E}_i) - \rho(\boldsymbol{X}_i;\ \boldsymbol{P}_{i,j}\boldsymbol{E}_j)|\boldsymbol{\mathcal{X}}] = 0\end{aligned}$$

Therefore, under the null, the statistics $G_i$ are statistically independent conditional on $\boldsymbol{\mathcal{X}}$, with expectations that sum to zero. They are marginally exchangeable by independence of experiments, and are thus marginally identically distributed. Nevertheless, they are not necessarily either conditionally or marginally Gaussian. To test the null hypothesis in such circumstances, a t-test that the mean of $G_i$ is zero seems to be a reasonable approximate approach, provided the distributions of the $G_i$ are close to symmetric [12,13]; this can be checked by usual methods such as histograms or QQ plots.

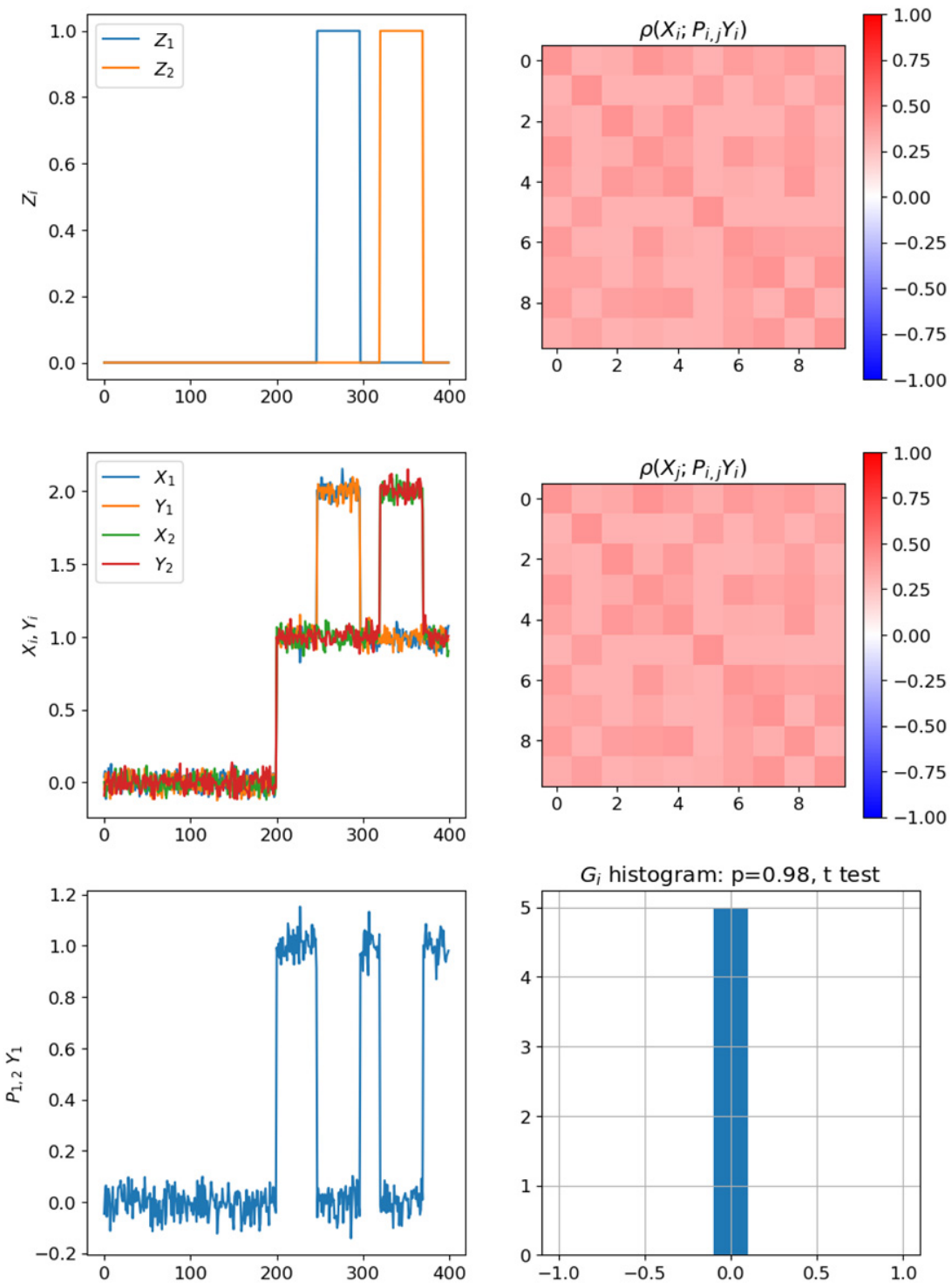

**Figure 2: correlations due to common effects of $Z$ are not detected.** $\boldsymbol{X}_i$ and $\boldsymbol{Y}_i$ are as in Fig. 1, but now $\boldsymbol{Z}_i$ is the step function that varies between experiments. Because both $\boldsymbol{Z}_1$ and $\boldsymbol{Z}_2$ are projected out, the similarity of $\boldsymbol{P}_{1,2}\boldsymbol{Y}_1$ to $\boldsymbol{X}_1$ is the same as to $\boldsymbol{X}_2$, so no correlation is detected.

## An illustration of Test 1

We illustrate this test with a simple example (Figs. 1, 2) in which $\boldsymbol{X}$ and $\boldsymbol{Y}$ are nearly identical, and are given by a sum of two step functions, plus independent additive noise. The first step function $\boldsymbol{S}_0$ is the same for each observation $i$, being 0 for the first half of the timeseries and 1 for the second half. The second step function $\boldsymbol{S}_{\mathrm{i}}$ is a short pulse which occurs at a different time for different experiments $i$, but always the same time for $\boldsymbol{X}$ and $\boldsymbol{Y}$. For each $i$ we simulate $\boldsymbol{X}_i = \boldsymbol{S}_0 + \boldsymbol{S}_i + \boldsymbol{\epsilon}_i^{(\boldsymbol{X})}$ and $\boldsymbol{Y}_i = \boldsymbol{S}_0 + \boldsymbol{S}_i + \boldsymbol{\epsilon}_i^{(\boldsymbol{Y})}$, where $\boldsymbol{\epsilon}_i^{(\boldsymbol{X})}$ and $\boldsymbol{\epsilon}_i^{(\boldsymbol{Y})}$ are both iid sequences of normal random variables with mean of zero and standard deviation of 0.05. We consider two cases: that $\boldsymbol{Z_i} = \boldsymbol{S_0}$ for all $i$, or that $\boldsymbol{Z_i} = \boldsymbol{S_i}$.

If $\boldsymbol{Z}_i = \boldsymbol{S}_0$, there is a partial correlation of $\boldsymbol{X}_i$ and $\boldsymbol{Y}_i$ given $\boldsymbol{Z}_i$, because the time of the pulse varies between experiments, but $\boldsymbol{Z}_i$ is always the same. Thus, $\rho(\boldsymbol{X}_i;\ \boldsymbol{P}_{i,j}\boldsymbol{Y}_i)$ is reliably larger than $\rho(\boldsymbol{X}_j;\ \boldsymbol{P}_{i,j}\boldsymbol{Y}_i)$, and the test finds statistical significance (Figure 1).

However, if $\boldsymbol{Z}_i = \boldsymbol{S}_i$, there is not a partial correlation of $\boldsymbol{X}_i$ and $\boldsymbol{Y}_i$ given $\boldsymbol{Z}_i$, since the correlation between $\boldsymbol{X}_i$ and $\boldsymbol{Y}_i$ can be explained by a common dependence on $\boldsymbol{Z}_i$. Thus, $\rho(\boldsymbol{X}_i;\ \boldsymbol{P}_{i,j}\boldsymbol{Y}_i)$ and $\rho(\boldsymbol{X}_j;\ \boldsymbol{P}_{i,j}\boldsymbol{Y}_i)$ are comparable, and the test finds no statistical significance (Figure 2).

To demonstrate that it is essential to use $\boldsymbol{P}_{i,j}\boldsymbol{Y}_i$ rather than simply $\boldsymbol{P}_i\boldsymbol{Y}_i$, we repeat the simulation with $\boldsymbol{Z}_i = \boldsymbol{S}_i$, but comparing $\rho(\boldsymbol{X}_i;\ \boldsymbol{P}_i\boldsymbol{Y}_i)$ with $\rho(\boldsymbol{X}_j;\ \boldsymbol{P}_i\boldsymbol{Y}_i)$. This invalid test erroneously finds a significant partial correlation where none exists (Figure 3).

## Test 2: An exact test using projection and permutation

A random permutation approach may be used for a test that provably controls the type 1 error rate. Let $h$ be a function that permutes the order of a sequence. For example, if $a = (7,8,9)$, we might have $h(a) = (8,9,7)$. We will use the notation $h(a)_i$ to refer to the $i$th element of $h(a)$, so that $h(a)_1 = 8$ in the example.

Let $\boldsymbol{\mathcal{Y}}$ be the sequence of all $\boldsymbol{Y}_i$. Similarly, let $\boldsymbol{\mathcal{E}} = (\boldsymbol{E}_1, \cdots, \boldsymbol{E}_n)$ be the sequence of all $\boldsymbol{E}_i$, and recall that $\boldsymbol{\mathcal{X}}$ is the sequence of all pairs $(\boldsymbol{X}_i, \boldsymbol{Z}_i)$. Let $\boldsymbol{P} = \boldsymbol{I} - \mathcal{M}([\boldsymbol{Z}_1\, \boldsymbol{Z}_2 \cdots \boldsymbol{Z}_n])$ where $[\boldsymbol{Z}_1\, \boldsymbol{Z}_2 \cdots \boldsymbol{Z}_n]$ is a block matrix, meaning that $\boldsymbol{P}$ is the $T \times T$ projection matrix orthogonal to the columns of all the $\boldsymbol{Z}_i$. We define the statistic

$$T(\boldsymbol{\mathcal{X}}, \boldsymbol{\mathcal{Y}}) = \frac{1}{n}\sum_{i=1}^{n} \rho(\boldsymbol{P}\boldsymbol{X}_i, \boldsymbol{P}\boldsymbol{Y}_i)$$

Under the null hypothesis, $\boldsymbol{Y}_i = \boldsymbol{Z}_i\boldsymbol{W}_i + \boldsymbol{E}_i$, so $\boldsymbol{P}\boldsymbol{Y}_i = \boldsymbol{P}\boldsymbol{E}_i$ and

$$T(\mathcal{X},\mathcal{Y}) = T(\mathcal{X},\mathcal{E}) = \frac{1}{n}\sum_{i=1}^{n} \rho(\boldsymbol{P}\boldsymbol{X}_i, \boldsymbol{P}\boldsymbol{E}_i)$$

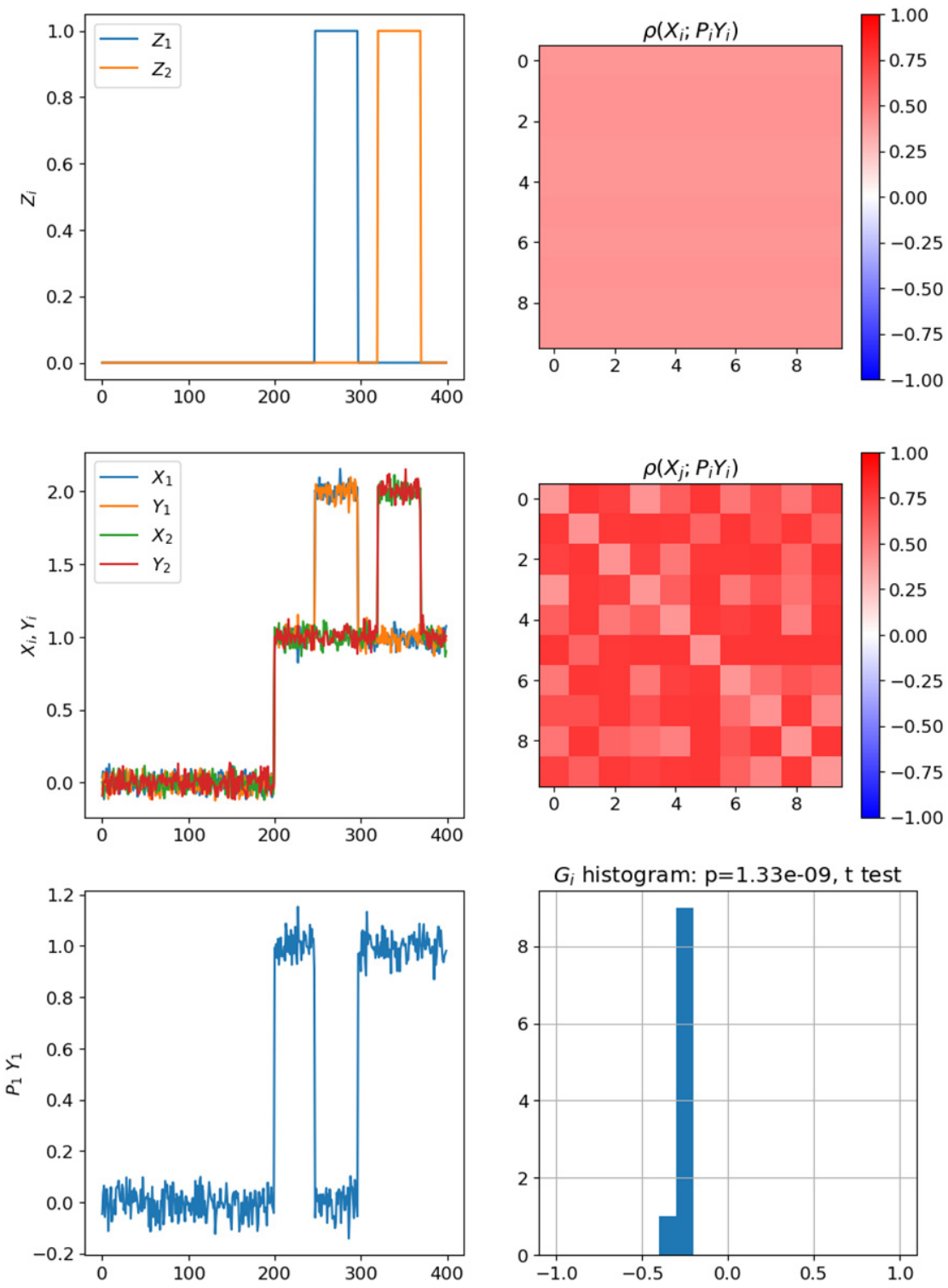

**Figure 3: both $Z_i$ and $Z_j$ must be projected out to avoid erroneous correlations.** All data are as in Fig. 2 but only $Z_i$ is projected from $Y_i$. Because $Z_j$ is not projected from $Y_i$, the residual $P_iY_i$ is more similar to $X_j$ than $X_i$, leading to an erroneously significant negative partial correlation.

To test the null we may compare the above statistic with an ensemble consisting of values of the form

$$T\big(\mathcal{X}, h(\mathcal{Y})\big) = \frac{1}{n}\sum_{i=1}^{n} \rho(\boldsymbol{P}\boldsymbol{X}_i, \boldsymbol{P}\boldsymbol{Y}_{h(i)})$$

in which each $\boldsymbol{X}_i$ is compared to a $\boldsymbol{Y}$ from a session chosen randomly without replacement. Because $\boldsymbol{P}$ is orthogonal to all the $Z_i$, under the null

$$T\big(\mathcal{X}, h(\mathcal{Y})\big) = T\big(\mathcal{X}, h(\mathcal{E})\big) = \frac{1}{n}\sum_{i=1}^{n} \rho(\boldsymbol{P}\boldsymbol{X}_i, \boldsymbol{P}\boldsymbol{E}_{h(i)})$$

Also under the null, all $\boldsymbol{E}_i$ are independent of each other and of $\mathcal{X}$. Thus, the distribution of the pair $\big(\mathcal{X}, h(\mathcal{E})\big)$ is the same for all permutations $h$, including the identity permutation.

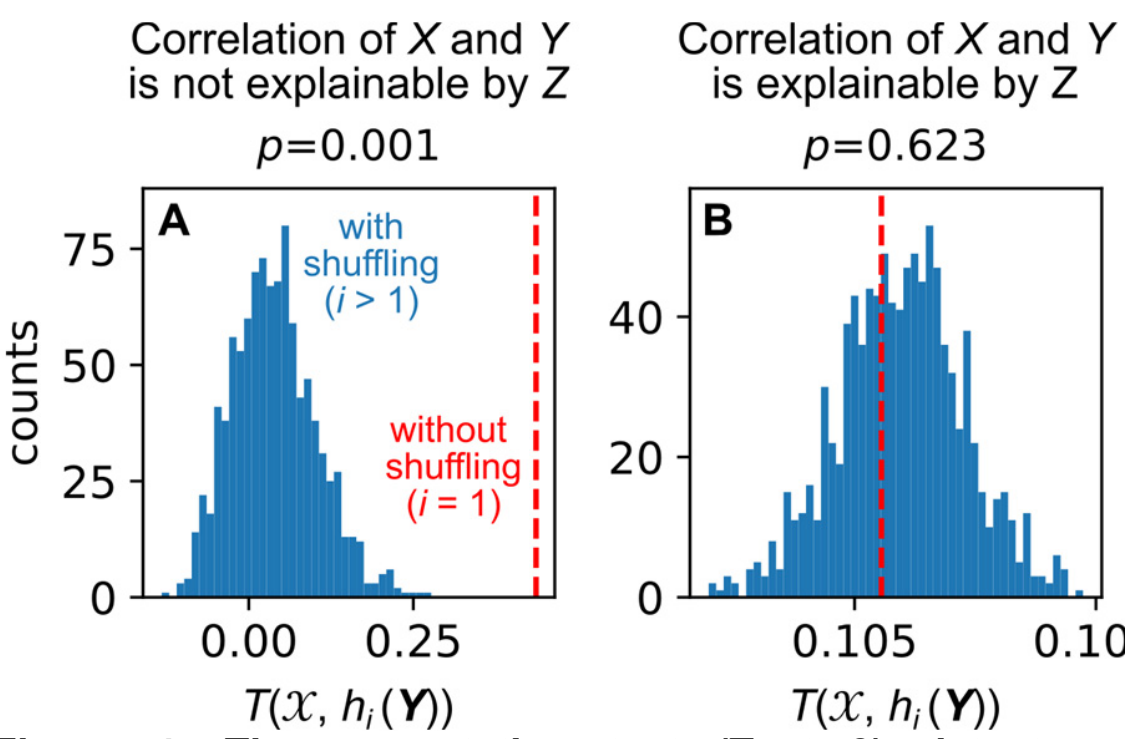

**Figure 4: The permutation test (Test 2) detects a correlation that cannot be explained by $Z$, but not a correlation that can be explained by $Z$.** In panels A and B respectively, the test was applied to the same data as in Figs. 1 and 2. We used $m - 1 = 999$ random shuffles.

We can therefore obtain a valid permutation test by comparing $T(\mathcal{X},\mathcal{Y})$ to a null ensemble $T\big(\mathcal{X}, h(\mathcal{Y})\big)$ with randomly chosen permutations $h$. The proof of this follows the standard proof for the validity of permutation tests[14,15]. Let $\mathcal{O}(\mathcal{E})$ be the orbit of $\mathcal{E}$ under the set of all permutations of the $n$ experiments. Then the probability $\mathbb{P}[\mathcal{E}|\mathcal{X}, \mathcal{O}(\mathcal{E})]$ is equal for each member of $\mathcal{O}(\mathcal{E})$**.** Thus if we let $h_1 \dots h_m$ be independent random permutations, then the $m+1$ random variables $T(\mathcal{X},\mathcal{Y}) = T(\mathcal{X},\mathcal{E})$ and $T\big(\mathcal{X}, h_l(\mathcal{Y})\big) = T\big(\mathcal{X}, h_l(\mathcal{E})\big)$, $l = 1 \dots m$, are independent and identically distributed on $(\mathcal{X}, \mathcal{O}(\mathcal{E}))$. If we let $R$ be the rank of $T(\mathcal{X},\mathcal{Y})$ within these numbers, breaking any ties randomly, then $R$ is a random integer uniformly distributed between 1 and $m+1$, and

$$p = \frac{R}{m+1}$$

is an exact p-value, obeying $\mathbb{P}(p \le \alpha) = \alpha$ for any $\alpha$ in the set of attainable p-values, i.e. $\{1/(m+1), 2/(m+1), \dots, 1\}$.

## An illustration of Test 2

In Fig. 4 we illustrate Test 2 (the permutation test just now described) by applying it to the same example data as shown in Figs. 1 and 2. The results are similar to those obtained with Test 1, with Test 2 detecting the correlation that cannot be explained by $\boldsymbol{Z}$, and appropriately failing to detect the correlation that is explained by $\boldsymbol{Z}$.

## Discussion

We have described two families of statistical tests for partial correlation between vectorial timeseries. The timeseries are not assumed stationary or linear, and in fact the tests apply to any data consisting of vectors depending on an index $t$. The tests require multiple observations of these timeseries. Intuitively, the tests measure how much better one can predict $\boldsymbol{Y}_i$ from $\boldsymbol{X}_i$ measured in the same experiment, compared to predicting $\boldsymbol{Y}_i$ from $\boldsymbol{X}_j$ measured in a different experiment. To remove a possible common effect of the confounding variable $\boldsymbol{Z}$, the first test projects out both $\boldsymbol{Z}_i$ and $\boldsymbol{Z}_j$ from the T-dimensional vector $\boldsymbol{Y}_i$. The second test projects out the all the matrices $\boldsymbol{Z}_1 \dots \boldsymbol{Z}_n$; this potentially reduces the power of the test but allows a proof that the p-value is valid even for small numbers of experiments.

The tests depend on a user-supplied measure $\rho(\boldsymbol{X}, \boldsymbol{Y})$ of the degree of association between two vectorial timeseries. They make no assumptions about this measure, so anything can be used. When dealing with 1-dimensional timeseries (as in the examples presented here), Pearson correlation is a natural choice. In higher dimensions one could for example use the fraction of variance explained by multiple linear regression or ridge regression. If $\boldsymbol{Y}$ is high dimensional, it may be advisable to use regularization or methods such as reduced rank regression or canonical correlation analysis. Using cross-validation to assess performance might improve statistical power, but this is not required for a valid test.

The tests may find use in multiple scientific fields where time-dependent experiments are performed repeatedly. An example in neuroscience would be if one makes multivariate recordings from the brains of subjects performing a behavioral task, and would like to test if some behavioral variables correlate with brain activity even after the effect of other behavioral variables are controlled for. Unlike the "pseudosession method" [10], the current tests require multiple repeats of an experiment. The matrices $\boldsymbol{W}_i$ can differ between experiments, and so it is possible for these experiments to be made from different subjects, or even contain vectors of different dimensions (e.g. recordings of different numbers of neurons). However to apply the test, all timeseries must have the same number of time points, which can be achieved by truncating them to have the same length if necessary. Furthermore, unlike the "pseudosession method" this method can be used when the predictor variables $\boldsymbol{X}$ and $\boldsymbol{Z}$ are not randomly generated, but produced by the subject themselves according to an unknown distribution.